# Tuning Electronic and Magnetic Properties of Early Transition Metal Dichalcogenides via Tensile Strain

Hongyan Guo[1,2,⊥], Ning Lu[3,2,⊥], Lu Wang[4,1], Xiaojun Wu[1,5,6]*, Xiao Cheng Zeng[2,5,]*

[1]CAS Key Lab of Materials for Energy Conversion, Department of Materials Science and Engineering, University of Science and Technology of China, Hefei, Anhui 230026, China, [2]Department of Chemistry and Department Mechanics and Materials Engineering, University of Nebraska-Lincoln, Lincoln, NE 68588, USA, [3]Department of Physics, Anhui Normal University, Wuhu, Anhui, 241000, China, [4]Department of Physics, University of Nebraska-Omaha, Omaha, NE 68182, USA, [5]Hefei National Laboratory for Physical Sciences at the Microscale, University of Science and Technology of China, Hefei, Anhui 230026, China, [6] Synergetic Innovation Center of Quantum Information & Quantum Physics, University of Science and Technology of China, Hefei, Anhui 230026, China

*Email: xzeng1@unl.edu; xjwu@ustc.edu.cn
[⊥]Both authors contribute equally to this work.

## Abstract

We have performed a systematic first-principles study of the effect of tensile strains on the electronic properties of early transition-metal dichalcogenide (TMDC) monolayers $MX_2$ (M = Sc, Ti, Zr, Hf, Ta, Cr; X = S, Se, and Te). Our density-functional theory (DFT) calculations suggest that the tensile strain can significantly affect the electronic properties of many early TMDCs in general and the electronic bandgap in particular. For group IVB TMDCs ($TiX_2$, $ZrX_2$, $HfX_2$), the bandgap increases with the tensile strain, but for $ZrX_2$ and $HfX_2$ (X=S, Se), the bandgap starts to decrease at strain 6% to 8%. For the group-VB TMDCs ($TaX_2$), the tensile strain can either induce the ferromagnetism or enhance the existing ferromagnetism. For the group-VIB TMDCs ($CrX_2$) the direct-to-indirect bandgap transition is seen upon application of the tensile strain, except $CrTe_2$ whose bandgap decreases with the tensile strain even though the direct character of its bandgap is retained. Lastly, for the group-IIIB TMDCs ($ScX_2$) in the T metallic phase, we find that the tensile strain has little effect on their electronic and magnetic properties. Our study suggests that strain engineering is an effective approach to modify electronic and magnetic properties of most early TMDC monolayers, thereby opening an alternative way for future optoelectronic and spintronic applications.

## I. Introduction

Two-dimensional (2D) monolayer materials with one-atom thickness, such as graphene, boron-nitride and ZnO, have aroused great interests due to their unique properties not seen in their bulk counterparts.[1-4] Recently, another class of 2D materials with three-atom thickness, *i.e.* the transition-metal dichalcogenides (TMDCs) such as $MoS_2$ and $WS_2$[5-8], has received intensive attention not only because of their novel electronic and catalytic properties but also owing to their wide-range tunability via strain or vertical electric field engineering. Indeed, engineering the bandgap of 2D materials via the tensile strain is one of the most viable approaches to achieve wide-range tunability in electronic properties for applications.[9-27] Previous first-principles calculations have shown that the monolayer and bilayer of $MoX_2$ and $WX_2$ (X = S, Se, and Te) can undergo the direct-to-indirect bandgap transition under increasing tensile strains.[14, 15, 19] Later experiments confirm the strain modulation of bandgap for both monolayer and bilayer $MoS_2$.[20, 24-26] Besides the bandgap, previous studies have also shown that the magnetic moments of $VS_2$ and $VSe_2$ monolayers can be enhanced rapidly with increasing the tensile strain[16]. Moreover, the nonmagnetic $NbS_2$ and $NbSe_2$ monolayers can turn into ferromagnetic under the tensile strain.[17]

The *d*-electron count of the transition metal is crucial to the TMDC phase. In contrast to the group-VIB bulk TMDCs which exhibit primarily the 2H structure, all group-IVB TMDCs are 1T structure, while group-VB TMDCs can be either 2H or 1T structures. Interestingly, for the 2H and 1T bulk structures, the non-bonding *d* bands located within the bandgap of bonding and anti-bonding states exhibit different splitting patterns, and the progressive filling of the non-bonding *d* bands results in diverse properties for the group-VB TMDCs.[8]

Thus far, most studies of the strain engineering have been focused on the group-VB ($VX_2$ and $NbX_2$, X = S, Se) and group-VIB TMDCs ($MoX_2$ and $WX_2$, X = S, Se). Over the past three years, 2D TMDCs in other early transition-metal groups have been synthesized.[28-31] Moreover, it has been shown from theoretical calculations that a series of $MX_2$ possess formation energies close to the monolayer $MoS_2$ and $WS_2$, suggesting high possibility of mechanical exfoliation of a single-layer $MX_2$ from the corresponding bulk material.[32, 33] Thus, it is timely to investigate the strain effect on the electronic properties of these TMDC monolayers, ranging from group IIIB to VIB.

In this article, we report calculation results of strain engineering of the bandgap of a series of early TMDC monolayers (M= Zr, Hf, Ta, Ti, Cr, Sc, X=S, Se, Te). Our computational study is motivated by several experiments reported in 2013 on successful application of either the uniform uniaxial strain and/or uniform biaxial strain on $MoS_2$ flakes using either the three-point bending configuration or piezoelectric substrates.[20, 22, 24, 26] Hence, both biaxial and uniaxial tensile strains are considered in our study. In light of 2D materials such as $MoS_2$ can sustain strains greater than 11%, as demonstrated in a previous experiment,[34] we also examine tensile strains up to 10% in our calculations. In most cases, the tensile strains can notably change the bandgap of early TMDC monolayers. Some generic trends on the bandgap engineering by strain are discussed.

## II. Computational Methods

All calculations are performed within the framework of spin-polarized density functional theory (DFT), implemented in the Vienna ab initio simulation package (VASP 5.3).[35, 36] The generalized gradient approximation (GGA) with the Perdew-Burke-Ernzerhof (PBE) functional and projector augmented wave (PAW) potentials are used.[37-39] Specifically, we adopt a $1 \times 1$ unit cell for structural optimization except $TaX_2$ for which a $2 \times 2$ supercell is used to investigate the magnetic properties. The vacuum size between two adjacent images is larger than 15 Å. An energy cutoff of 500 eV is adopted for the plane-wave expansion of the electronic wave function. Geometry structures are relaxed until the force on each atom is less than 0.01 eV/Å and the energy convergence criteria of $1 \times 10^{-5}$ eV are met. The 2D Brillouin zone integration using the $\Gamma$-center scheme is sampled with $9 \times 9 \times 1$ grid for geometry optimizations and $15 \times 15 \times 1$ grid for static electronic structure calculations. For each monolayer system, the unit cell is optimized to obtain the lattice parameters corresponding to the lowest total energy. The tensile strain is undertaken in three different ways: uniaxial expansion of the monolayer in $x$-direction (xx), $y$-direction (yy), or homogeneous biaxial expansion in both $x$- and $y$-directions (xy). The strain is defined as $\varepsilon=\Delta m/m_0$. For biaxial tensile strain, $m_0$ is unstrained cell parameters and $\Delta m+m_0$ is strained cell parameters. For uniaxial tensile strain, $m_0$ is the projection of unit-cell vectors in the $x$ or $y$ direction, and $\Delta m$ is the corresponding change of $m_0$. Tensile strain ranging from 0 to 10% is considered. Lastly, we note that for monolayer $MoS_2$, the measured bandgap is around 1.8 eV,[20] while our PBE calculation gives a bandgap about 1.67 eV. Note also that the calculation based on a hybrid density functional (HSE06) gives a bandgap about 2.23 eV and the $G_0W_0$ calculation gives a bandgap about 2.78 eV, both overestimating the experimental bandgap.[40] Thus, in this study, the calculated bandgaps of 2D TMDCs are all based on the PBE calculations.

## III. Results and Discussion

To investigate the strain effect on the electronic properties of early TMDCs, we first relax the atomic positions and lattice vectors to obtain the optimized geometry of each monolayer $MX_2$ (M= Sc, Ti, Zr, Hf, Ta and Cr; X=S, Se, and Te). Although various stable and metastable bulk phases of TMDCs have been reported in previous studies,[41-45] in this study, we only consider the most stable phase of TMDCs monolayers at 0 K.[32, 33, 46] For $ScX_2$, $TiX_2$, $ZrX_2$, $HfX_2$, the most stable phase is the T phase; and for $TaX_2$, $CrX_2$, the most stable phase is the H phase. The optimized value of lattice constant $a_0$ is listed in Table 1. The strained cell is modeled by stretching the hexagonal ring in the $x$- or/and $y$-directions, as shown in Figure 1. The irreducible path $\Gamma KM\Gamma$ corresponds to the irreducible Brillouin zone of unstrained and symmetrically strained ($xy$) TMDCs, while $\Gamma KMLM'\Gamma$ encloses the irreducible wedge for the asymmetrically strained ($xx$ or $yy$) TMDCs.

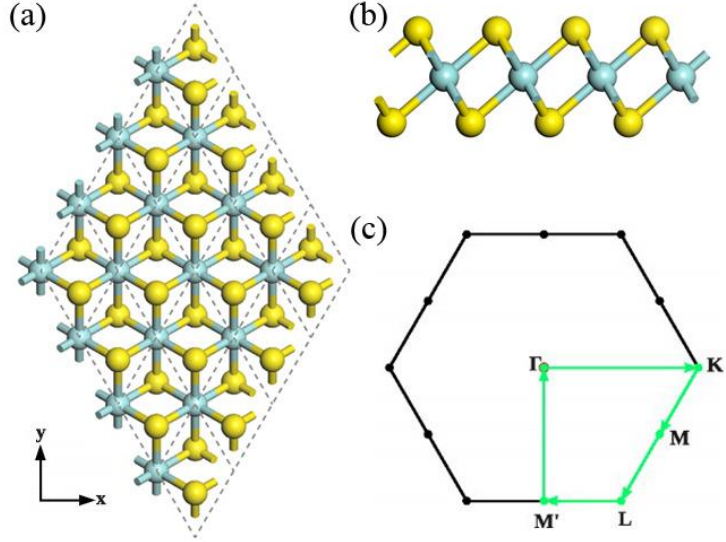

**Figure 1**. (a) Top and (b) side view of monolayers MX$_2$ in the T phase, where blue and yellow balls represent M and X atoms, respectively. The figure shows a $4 \times 4$ supercell. The strain along $x$ or/and $y$-directions is applied by varying the $x$ or/and $y$ components of these lattice vectors, respectively. (c) Irreducible Brillouin zone of MX$_2$. The irreducible path ΓKMΓ corresponds to the unstrained and symmetrically strained TMDCs, while ΓKMLM'Γ encloses the irreducible wedge for the asymmetrically strained TMDCs.

**Table 1**. The optimized lattice constant a$_0$ (in Å) of MX$_2$ monolayers (M= Sc, Ti, Zr, Hf, Ta, and Cr; X=S, Se, and Te).

|          | ScX$_2$ | TiX$_2$ | ZrX$_2$ | HfX$_2$ | TaX$_2$ | CrX$_2$ |
|----------|---------|---------|---------|---------|---------|---------|
| **X=S**  | 3.77    | 3.39    | 3.68    | 3.64    | 3.34    | 3.05    |
| **X=Se** | 3.63    | 3.53    | 3.80    | 3.76    | 3.46    | 3.22    |
| **X=Te** | 3.83    | 3.74    | 3.98    | 3.97    | 3.69    | 3.41    |

Next, the band structures of unstrained monolayer MX$_2$ are computed. DFT (PBE) calculations show that ZrS$_2$, ZrSe$_2$, HfS$_2$ and HfSe$_2$ monolayers are semiconductors with an indirect bandgap, while CrX$_2$ (X=S, Se, and Te) are semiconductors with a direct bandgap; $ZrTe_2$, $HfTe_2$, as well as $ScX_2$, $TiX_2$, and $TaX_2$ (X=S, Se, and Te) are all metals. The computed bandgap of $ZrS_2$, $ZrSe_2$, $HfS_2$, $HfSe_2$, $CrS_2$, $CrSe_2$ and $CrTe_2$ is 1.10, 0.45, 1.27, 0.61, 0.92, 0.74, and 0.60 eV, respectively. These values are in agreement with previous calculations.[32, 33]

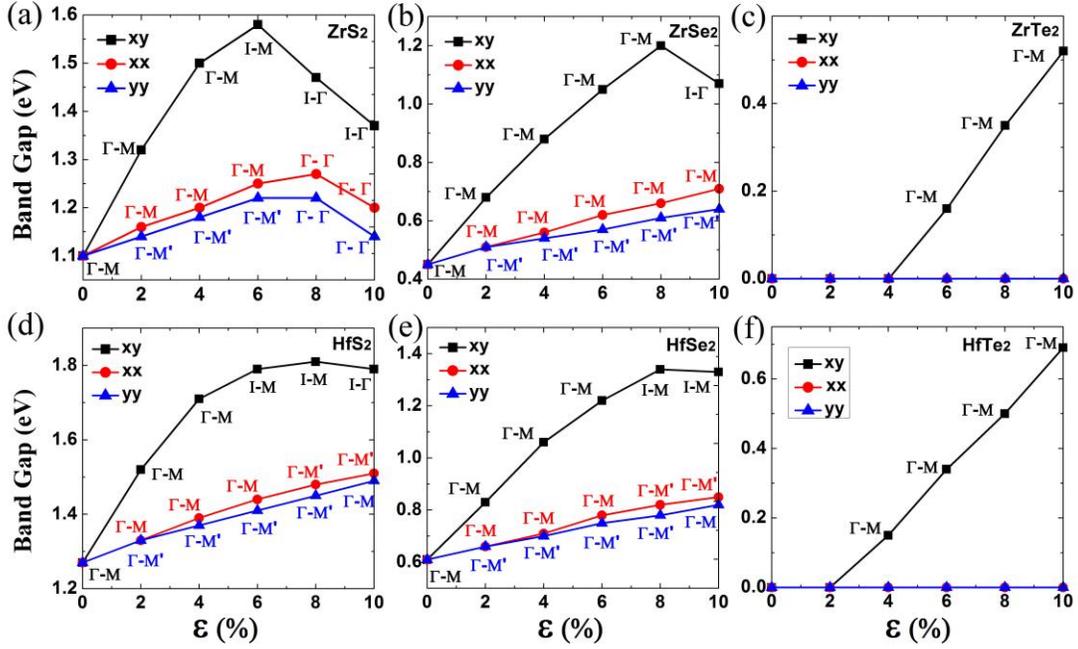

**Figure 2.** Computed bandgaps (PBE) of (a) – (c) $ZrX_2$ (X=S, Se, Te) and (b)-(d) $HfX_2$ (X=S, Se, Te) monolayers as a function of strain ranging from 0 to 10%. Strain is applied to the optimized structures ($\varepsilon$ =0) through three different approaches: uniaxial expansion in $x$-direction ($xx$), $y$-direction ($yy$), or homogeneous expansion in both $x$- and $y$-directions ($xy$).

## A. $ZrX_2$, $HfX_2$ and $TiX_2$

Effect of tensile strain on the electronic structures of monolayers $MX_2$ (M=Ti, Zr and Hf; X=S, Se and Te) is then studied. Figure 2(a)-(f) shows the computed bandgaps of monolayer $MX_2$ (M =Zr, Hf; X = S, Se and Te) with respect to the tensile strains, respectively.

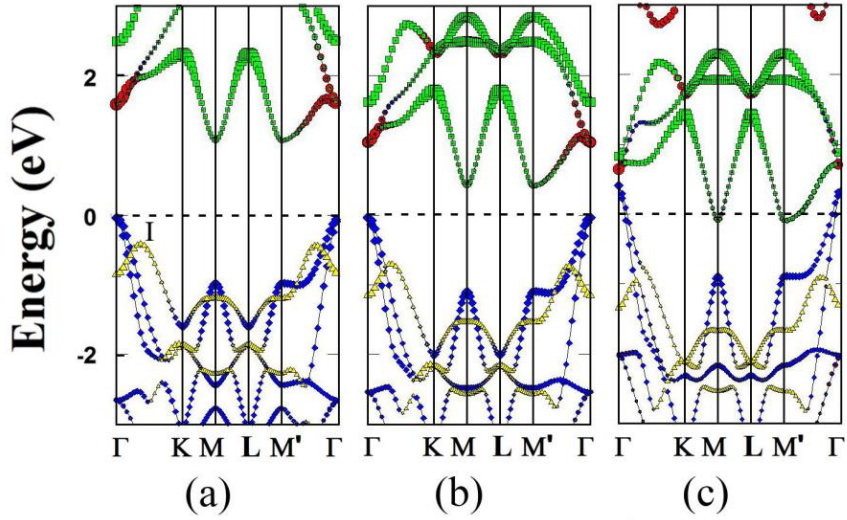

**Figure 3.** The computed band structures (PBE) of unstrained monolayer (a) ZrS$_2$, (b) ZrSe$_2$, and (c) ZrTe$_2$, respectively. The red line represents $d_{xy}$, $d_{yz}$, and $d_{xz}$ orbitals of Zr atom, the green line represents $d_z^2$ and $d_{x^2-y^2}$ orbitals of Zr atom, the blue line represents $p_x$ and $p_y$ orbitals of (a) S, (b) Se, and (c) Te atom, and the yellow line represents $p_z$ orbital of (a) S, (b) Se, and (c) Te atom. The dashed line represents the Fermi level.

For ZrS$_2$, its bandgap increases with the strain initially and then decreases when the biaxial (uniaxial) tensile strain is beyond 6% (8%). Note that the unstrained monolayer ZrS$_2$ is a semiconductor with an indirect bandgap of 1.10 eV (Figure 3(a)). Here, the valence-band maximum (VBM) and conduction-band minimum (CBM) are located at the Γ and M point, respectively. The VBM is mainly contributed by $p_x$ and $p_y$ orbitals of S atom, and the I point (Figure 3(a)) located on the Γ-K line is mainly contributed by $p_z$ orbital of S atom. The CBM is mainly contributed by the $d_z^2$ and $d_{x^2-y^2}$ orbitals of Zr atom, and the Γ point of the conduction band is mainly contributed by the $d_{xy}$, $d_{yz}$, and $d_{xz}$ orbitals of Zr atom.

Upon applying a 2% biaxial tensile strain in both $x$- and $y$-directions (xy), the bandgap of ZrS$_2$ increases to 1.32 eV while the VBM and CBM are still located at the Γ and M point, respectively. When the biaxial tensile strain increases to 6% (Figure 4), the CBM is still located at the M point whereas the VBM is relocated to I point, and the bandgap increases to 1.58 eV. Here, the energy of the $p_z$ orbital of S atom at I point is higher than that of the $p_x$ and $p_y$ orbitals of S atom at Γ point, resulting in the relocation of the VBM from $\Gamma$ to I point. When the strain further increases to 8%, the energy of the $d_{xy}$, $d_{yz}$, and $d_{xz}$ orbitals of Zr atom at the $\Gamma$ point is lower than that of the $d_z^2$ and $d_{x^2-y^2}$ orbitals of Zr atom at the M point, results in the relocation of CBM to the $\Gamma$ point. As such, the bandgap decreases to 1.47 eV.

More importantly, in the case of uniaxial tensile strain along $x$- or $y$- direction, our calculation shows that ZrS$_2$ monolayer transforms from an indirect bandgap to a direct bandgap semiconductor when the uniaxial tensile strain reaches to 8% (see Figure 2(a) and Figure 4). The direct bandgap character is still kept when the strain increases to 10%. Under the uniaxial tensile strain, the energy of the $d_z^2$ and $d_{x^2-y^2}$ orbitals of Zr atom at M point shifts upward, eventually above the energy of the $d_{xy}$, $d_{yz}$, and d$_{xz}$ orbitals of Zr atom at $\Gamma$ point. Hence, the CBM shifts from M point to $\Gamma$ point, while the VBM at the $\Gamma$ point is not affected, leading to the indirect-to-direct bandgap transition.

Similar to ZrS$_2$, the unstrained monolayer ZrSe$_2$ has similar state contributions around the Fermi level (see Figure 3(b)). As shown in Figure 2(b)), upon applying the biaxial tensile strain from 0 to 8% to ZrSe$_2$, the bandgap increases from 0.45 to 1.20 eV. Under 10% strain, however, the bandgap decreases to 1.07 eV. Upon applying the uniaxial tensile strain, the bandgap increases with increasing the strain. However, the indirect bandgap character is unchanged. Note also that HfS$_2$ and HfSe$_2$ exhibit similar strain-dependent bandgap behavior as ZrSe$_2$ (see Figure 2(d)-(e), Supporting Information Figures S1 and S2).

Unlike ZrS$_2$, unstrained ZrTe$_2$ and HfTe$_2$ monolayers are metallic. We find that the biaxial tensile strain (xy) leads to a metal-to-semiconductor transition (see Figure 2(c) and (f)). Upon the uniaxial tensile strain ($xx$ or $yy$), however, both ZrTe$_2$ and HfTe$_2$ maintain the metallic character. As shown in Figure 3(c), the valance band around the Fermi level is mainly contributed by $p$ orbital of Te atom, while the conduction band near the Fermi level is mainly contributed by $d$ orbital of Zr atom. The energy splitting between the $p$ orbital of Te atom and the $d$ orbital of Zr atom increases with increasing the biaxial tensile strain, leading to the metal-to-semiconductor transition.

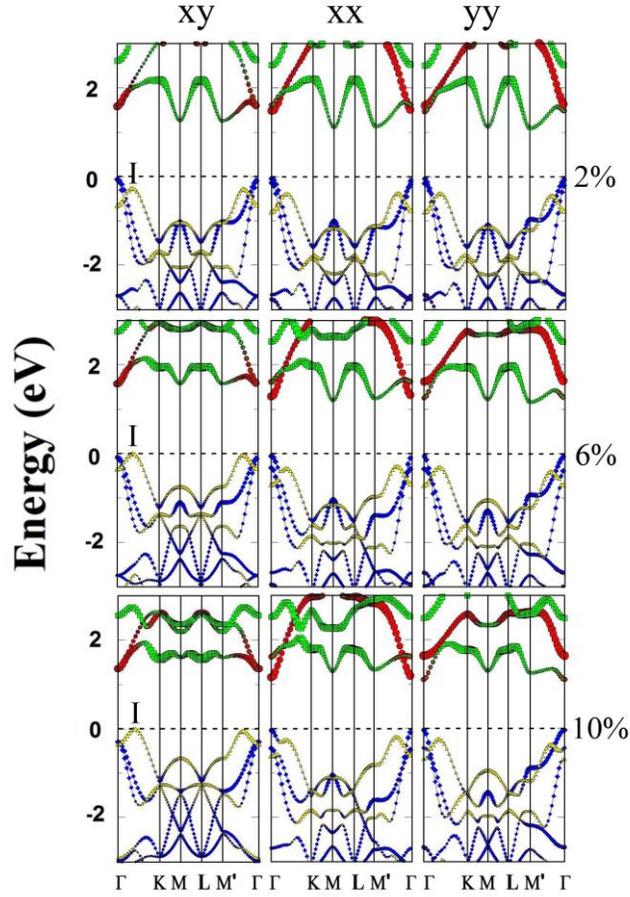

**Figure 4.** Computed band structures of monolayer $ZrS_2$ with respect to the biaxial tensile strain (*xy*), or uniaxial tensile strain in *x*-direction (*xx*) or *y*-direction (*yy*). The applied strain $\varepsilon$ = 2%, 6% and 10%, respectively. The red line denotes $d_{xy}$, $d_{yz}$, and $d_{xz}$ orbitals of Zr atom, the green line denotes $d_z^2$ and $d_{x^2-y^2}$ orbitals of Zr atom, the blue line denotes $p_x$ and $p_y$ orbitals of S atom, and the yellow line denotes $p_z$ orbital of S atom. The dashed line represents the Fermi level.

We also investigate the effect of the tensile strain on the bandgap of $TiX_2$ (X = S, Se, and Te) monolayers. For $TiS_2$, as shown in Figure 5(a), a metal-to-semiconductor transition occurs at a relatively low biaxial tensile strain of 2%. The bandgap increases to 0.56 eV when the biaxial tensile strain reaches to 10%. However, under the uniaxial tensile strain along *x*- or *y*-direction, the metal-to-semiconductor transition takes place at a relatively large uniaxial strain. In contrast, both $TiSe_2$ and $TiTe_2$ still retain the metallic character even under the largest tensile strain (10%) considered in this study.

**B. CrX$_2$**

As shown in Figure 5, the bandgap of CrX$_2$ (X=S, Se, and Te) monolayers decreases with increasing the tensile strain, showing a similar trend as other group VIB TMDCs such as MoS$_2$. Note that all unstrained CrX$_2$ monolayers (X=S, Se, and Te) are direct bandgap semiconductor (Figure 6). Taking unstrained CrS$_2$ as an example, the VBM and CBM are both located at K (L) point as shown in Figure 6(a). The valence band around the Γ point and K (L) point is mainly contributed by the $d_z{}^2$ orbital of Cr atom and the $d_{xy}$ and $d_{x^2-y^2}$ orbitals of Cr atom, respectively. The CBM at K (L) point is mainly contributed by the $d_z{}^2$ orbital of Cr atom. The tensile strain leads to a shift downward for the valence band around K (L) point, compared to the valence band around the Γ point. However, the CBM around the K and L points shows little shift under the tensile strain. Thus, a direct-to-indirect bandgap transition can occur even at 2% tensile strain. Moreover, the CBM around K and L points shifts downward to the Fermi level, leading to the bandgap decrease with increasing the tensile strain. The shift is more substantial in the case of biaxial tensile strain (xy), resulting in a semiconductor-to-metal transition when the biaxial tensile strain reaches to 10% (Figure 7).

For CrSe$_2$, the relatively low biaxial tensile strain (*xy*) of 2% can also induce an indirect bandgap, similar to the case of CrS$_2$ monolayer. However, under a 2% uniaxial tensile strain (*xx* or *yy*), CrSe$_2$ can still retain a direct bandgap. When the uniaxial tensile strain increases to 4%，CrSe$_2$ turns into an indirect bandgap semiconductor. Unlike CrS$_2$ and CrSe$_2$ monolayers, for CrTe$_2$, the direct-to-indirect bandgap transition only takes place under the biaxial tensile strain (*xy*) of 6%. In the cases of uniaxial tensile strain (*xx* or *yy*), the CrTe$_2$ monolayer still retains a direct bandgap, as shown in Figure 5(d).

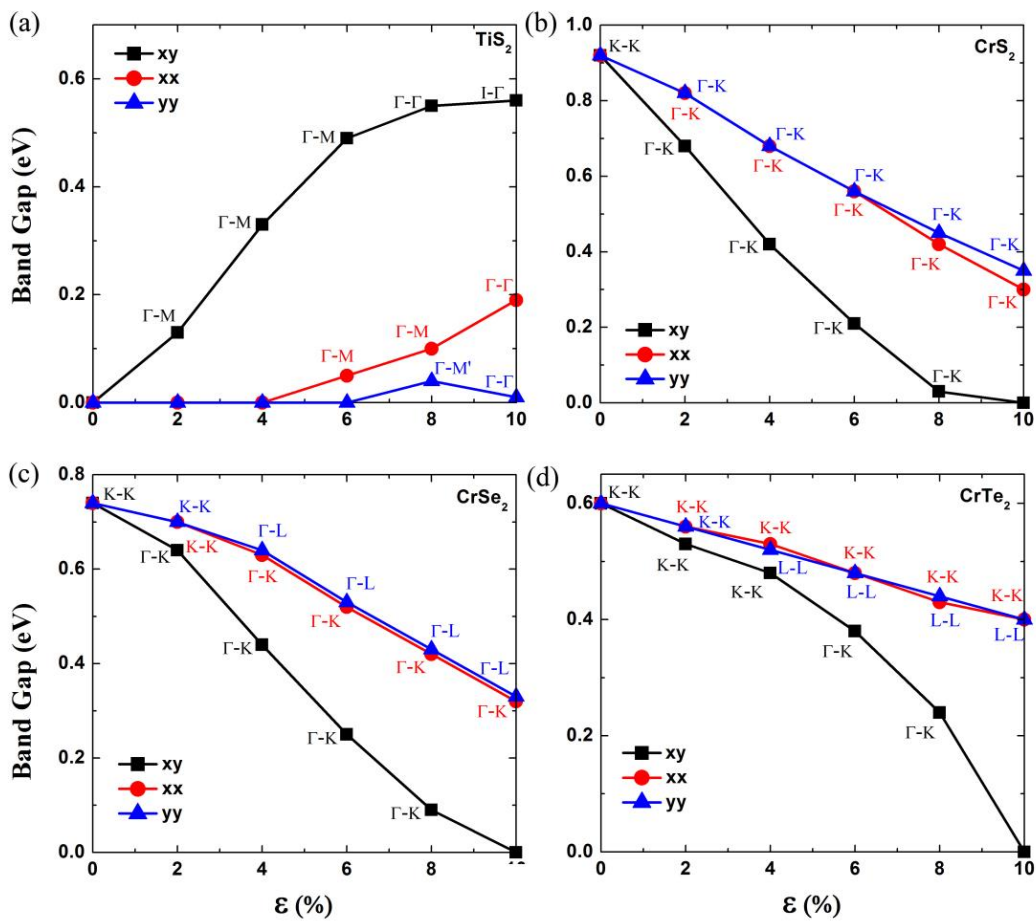

**Figure 5**. Bandgaps of MX$_2$ (M=Cr, Ti; X=S, Se, Te) monolayers with respect to the strain, ranging from 0 to 10%: (a) TiS$_2$ (b) CrS$_2$, (c) CrSe$_2$, and (d) CrTe$_2$.

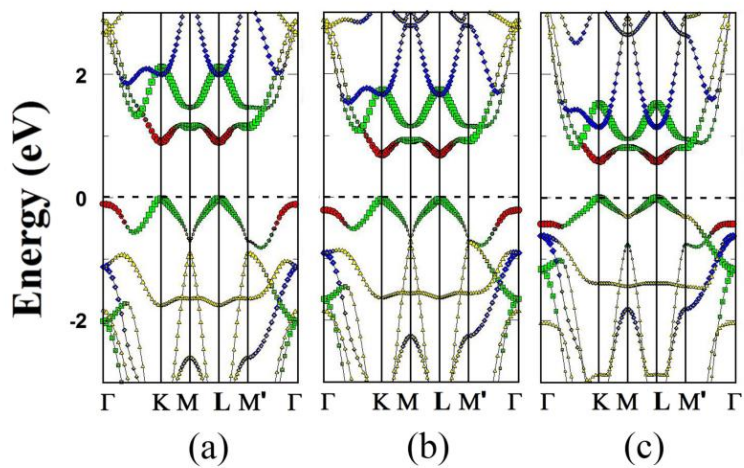

**Figure 6.** Computed band structures of unstrained monolayer: (a) CrS$_2$, (b) CrSe$_2$, and (c) CrTe$_2$. The red line denotes $d_z^2$ orbital of Cr atom, the green line denotes $d_{xy}$ and $d_{x^2-y^2}$ orbitals of Cr atom, the blue line denotes $d_{yz}$ and $d_{xz}$ orbitals of Cr atom, and the yellow line denotes $p$ orbital of (a) S, (b) Se, and (c) Te atom. The dashed line denotes the Fermi level.

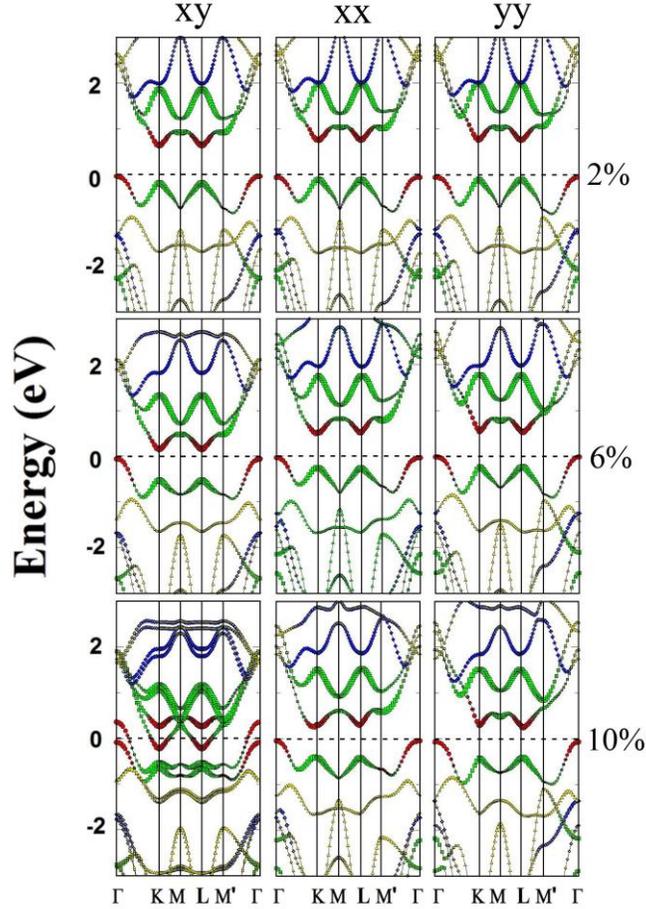

**Figure 7.** Computed band structures of CrS$_2$ monolayer with respect to the biaxial tensile strain (*xy*), or uniaxial tensile strain in *x*-direction (*xx*) or *y*-direction (*yy*). The applied strain ε = 2%, 6%, and 10%. The red line denotes $d_z^2$ orbital of Cr atom, the green line denotes $d_{xy}$ and $d_{x^2-y^2}$ orbitals of Cr atom, the blue line denotes $d_{yz}$ and $d_{xz}$ orbitals of Cr atom, and the yellow line denotes $p$ orbital of S atom. The dashed line denotes the Fermi level.

As we can see, for ZrX2 and HfX2, the states around the Fermi level are mainly contributed by M and X atoms respectively. The bonding length increase when the tensile strain increase, which will lead the reduction of the bonding interaction. And the energy difference between the

### C. TaX$_2$ and ScX$_2$

The group-IVB TMDC TaX$_2$ monolayers are all metallic. Both TaS$_2$ and TaSe$_2$ retain their metallic character under the tensile strain. For TaS$_2$, when the biaxial tensile strain reaches 8%, the degeneracy of spin-up and spin-down band structures no longer holds, leading to a nonmagnetic-to-ferromagnetic transition. The magnetic moment per supercell is mainly contributed by Ta atoms (1.88 $\mu_B$). At 10% tensile strain, the spin splitting becomes more significant and TaS$_2$ turns to a strong ferromagnetic metal with the magnetic moment of Ta atoms amounting to 1.98 $\mu_B$. Similar strain effect is also found for TaSe$_2$.

The unstrained TaTe$_2$ is a ferromagnetic metal. The total magnetic moment is mainly contributed by Ta atoms. Under a biaxial tensile strain, TaTe$_2$ still retains its ferromagnetic metallic character. Figure 8(a) shows the magnetic moment per Ta atom in the supercell as a function of the biaxial tensile strain. Clearly, the magnetic moment increases with increasing the biaxial tensile strain. The energy difference $\Delta E$ (=$E_{AFM}$ - $E_{FM}$) for the TaTe$_2$ as a function of biaxial tensile strain is shown in Figure 8(b). $\Delta E$ increases with increasing the biaxial tensile strain. The FM state is always more stable than the AFM state regardless of the biaxial tensile strain, which suggests that the FM coupling is stable in a wide range of biaxial tensile strains. As can be seen in Figure 9, the magnetism is mainly contributed from the Ta atoms. With increasing the tensile strain, the electronic spin charge for the Ta atoms increases, resulting in a higher magnetic moment on the Ta atoms. Similar strain-dependent magnetism is seen in the case of VX$_2$ monolayer.[16]

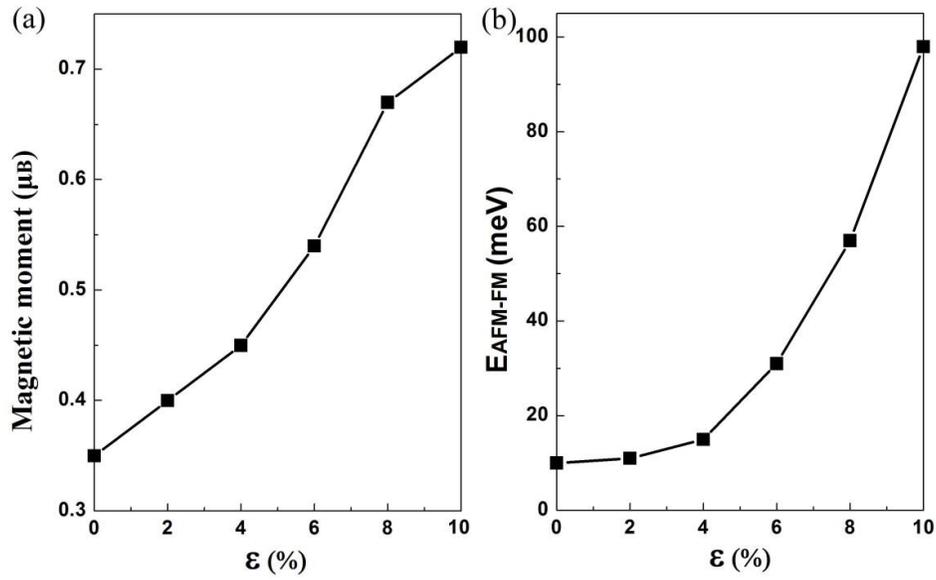

**Figure 8**. Biaxial tensile strain dependence of (a) magnetic moment per Ta atom in TaTe$_2$ supercell and (b) the energy difference per supercell between ferromagnetic (FM) and antiferromagnetic (AFM) coupling states.

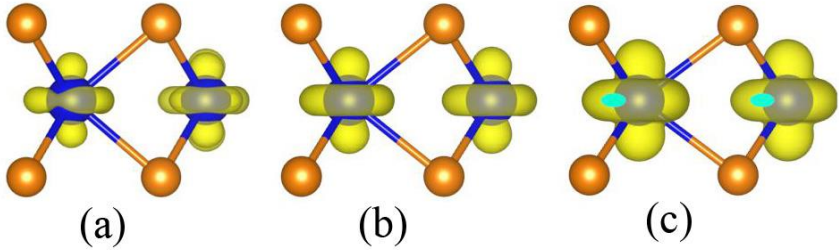

**Figure 9**. The spin charge densities of TaTe$_2$ with (a) 0% (b) 4% (c) 8% tensile strain. The iso-surface value is 0.006 e/bohr$^3$.

Lastly, for group-IIIB TMDCs ScX$_2$ (X=S, Se, and Te), the stable phase is T with metallic character. Unlike TMDCs in other groups, the tensile strain shows little effect on the electronic properties of ScX$_2$ while ScX$_2$ can retain its metallic character regardless of the strain (Figure 10).

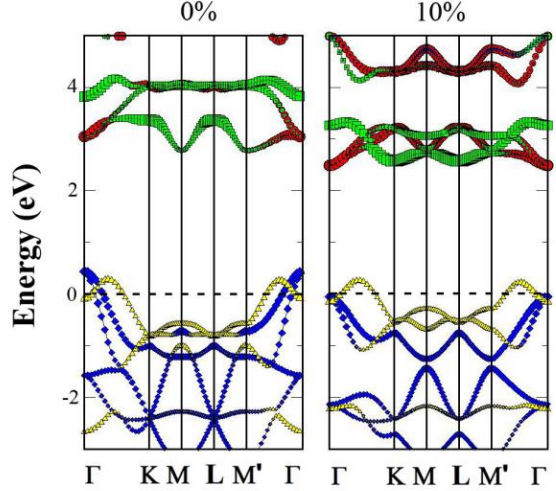

**Figure 10**. Computed band structures of unstrained ScS$_2$ monolayer (left) and strained ScS$_2$ (under a biaxial tensile strain of 10%) (right). The red line denotes $d_{xy}$, $d_{yz}$, and $d_{xz}$ orbitals of Sc atom, the green line denotes $d_z^2$ and $d_{x^2-y^2}$ orbitals of Sc atom, the blue line denotes $p_x$ and $p_y$ orbitals of S atom, and the yellow line denotes $p_z$ orbital of S atom. The dashed line represents the Fermi level.

## IV. Conclusion

We have performed a comprehensive study of the effect of tensile strains on the electric and magnetic properties of early TMDC monolayers. Our calculations show that the mechanical strain can play an important role in tuning the bandgap of most early TMDCs. Some generic trends of bandgap engineering due to the strain can be summarized for TMDCs in different groups. For group-IIIB TMDC ScX$_2$ with T metallic phase, the strain has little effect on their electronic properties. For group-VIB TMDC TiX$_2$, ZrX$_2$ and HfX$_2$, the bandgap increases with increasing the tensile strain initially but then decreases the bandgap beyond 6% to 8% strain for some ZrX$_2$ and HfX$_2$ monolayers, respectively. In particular, when the uniaxial tensile strain reaches 8%, ZrS$_2$ undergoes an indirect-to-direct bandgap transition. For group-V TMDC TaX$_2$, the strain can either induce or enhance ferromagnetism. Finally, for group-VIB TMDC CrX$_2$, the tensile strain can induce the direct-to-indirect bandgap transition except CrTe$_2$ which can retain its direct character regardless of the strain. In any case, the bandgap of CrX$_2$ decreases with the tensile strain. We expect the wide-range electronic and magnetic properties of early TMDC monolayers can be exploited for future optoelectronic and spintronic applications.

## Acknowledgements

XCZ is grateful to valuable discussions with Professors Ali Adibi, Eric Vogel, Joshua Robinson, and Ali Eftekhar. The USTC group is supported by the National Basic Research Programs of China (Nos. 2011CB921400, 2012CB 922001), NSFC (Grant Nos. 21121003, 11004180, 51172223), One Hundred Person Project of CAS, Strategic Priority Research Program of CAS (XDB01020300), Shanghai Supercomputer Center, and Hefei Supercomputer Center. UNL group is supported by ARL (Grant No. W911NF1020099), NSF (Grant No. DMR-0820521), Nebraska Center for Energy Sciences Research, and UNL Holland Computing Center, and by a grant from USTC for (1000plan) Qianren-B summer research.

**Supporting Information**. Computed band structures of $HfS_2$, $HfSe_2$, and $HfTe_2$ monolayers under different strains. This material is available free of charge via the Internet at http://pubs.acs.org.

**TOC Graphics**

| Strain Effect | No effect | Gap decrease | Gap Increase | Increase decrease | Magnetism |
|---|---|---|---|---|---|
| $ScS_2$ | $TiS_2$ | $ZrS_2$ | $HfS_2$ | $TaS_2$ | $CrS_2$ |
| $ScSe_2$ | $TiSe_2$ | $ZrSe_2$ | $HfSe_2$ | $TaSe_2$ | $CrSe_2$ |
| $ScTe_2$ | $TiTe_2$ | $ZrTe_2$ | $HfTe_2$ | $TaTe_2$ | $CrTe_2$ |